# SafeSpec: Banishing the Spectre of a Meltdown with Leakage-Free Speculation

Khaled N. Khasawneh
Department of Computer Science and Engineering
University of California, Riverside
Email: kkhas001@ucr.edu

Esmaeil Mohammadian Koruyeh
Department of Computer Science and Engineering
University of California, Riverside
Email: emoha004@ucr.edu

Chengyu Song
Department of Computer Science and Engineering
University of California, Riverside
Email: csong@cs.ucr.edu

Dmitry Evtyushkin
Department of Computer Science
College of William and Mary
Email: devtyushkin@wm.edu

Dmitry Ponomarev
Department of Computer Science
Binghamton University
Email: dima@cs.binghamton.edu

Nael Abu-Ghazaleh
Department of Computer Science and Engineering
University of California, Riverside
Email: naelag@ucr.edu

***Abstract*—Speculative execution, which is used pervasively in modern CPUs, can leave side effects in the processor caches and other structures even when the speculated instructions do not commit and their direct effect is not visible. The recent Meltdown and Spectre attacks have shown that this behavior can be exploited to expose privileged information accessed speculatively to an unprivileged attacker. In particular, the attack forces the speculative execution of a code gadget that will carry out the illegal read, which eventually gets squashed, but which leaves a side-channel trail that can be used by the attacker to infer the value. Several attack variations are possible, allowing arbitrary exposure of the full kernel memory to an unprivileged attacker. In this paper, we introduce a new model (*SafeSpec*) for supporting speculation in a way that is immune to the side-channel leakage necessary for attacks such as Meltdown and Spectre. In particular, *SafeSpec* stores side effects of speculation in separate structures while the instructions are speculative. The speculative state is then either committed to the main CPU structures if the branch commits, or squashed if it does not, making all direct side effects of speculative code invisible. The solution must also address the possibility of a covert channel from speculative instructions to committed instructions before these instructions are committed (i.e., while they share the speculative state). We show that *SafeSpec* prevents all three variants of Spectre and Meltdown, as well as new variants that we introduce. We also develop a cycle accurate model of modified design of an x86-64 processor and show that the performance impact is negligible (in fact a small performance improvement is achieved). We build prototypes of the hardware support in a hardware description language to show that the additional overhead is acceptable. *SafeSpec* completely closes this class of attacks, retaining the benefits of speculation, and is practical to implement.**

## I. Introduction

Speculative execution is a standard microarchitectural technique used in virtually all modern CPUs to improve performance. Recently, it has been shown that speculatively executed instructions can leave measurable side-effects in the processor caches and other shared structures even when the speculated instructions do not commit and their direct effect is not visible. The recent Meltdown and Spectre attacks [1], [2], [3], [4] (we call this class of attacks *speculation attacks*) have shown that this behavior can be exploited to expose information that is otherwise inaccessible. In a typical scenario, attackers either mis-train the branch predictor unit or directly pollute it [5] to force the speculative execution of code that reads privileged data (these checks are not enforced during speculation). Although the speculative instructions will eventually get squashed, leaving no direct data accessible to the attacker, they leave a side-channel trail that can be used to infer the value. Several attack variations have been demonstrated, including arbitrary exposure of the full memory of other processes, OS kernel, hypervisor, and even SGX enclaves [6] to an unprivileged attacker, making this a dangerous open attack vector on modern systems. We describe these attacks and present our threat model in Section II.

Although a number of defenses and software patches have been proposed to mitigate Spectre and Meltdown [7], [8], they often address only one aspect of the attack, leaving attackers with other possible variations that are still available. In addition, these patches often lead to high overheads: 10-30% reported on average, but often much higher. For example, Netflix reported 800% slowdown with the Meltdown patches on their systems [9], [10]. Most of the solutions target a subset of the threat models and make assumptions that can be broken by future architectures

In this paper, we explore whether speculation can be made leakage free in a principled way, enabling CPUs to retain the performance advantages of speculation while removing the security vulnerabilities that speculation exposes. To this end, we introduce *SafeSpec*, a design principle where speculative state is stored in temporary structures that are not accessible by committed instructions. As instructions transition from being speculative to commitable, any speculative state is moved to the permanent structures. On the other hand, if a speculative branch is squashed, the speculative side effects are canceled in place leaving no measurable side effects in the permanent structures and closing the vulnerability exploited by speculation attacks. We consider two variants that differ in when an instruction is considered safe to commit. In the wait-for-commit (WFC) variation, an instruction is considered speculative until it is committed. In the wait-for-branch (WFB) variation, we consider an instruction to be commitable when the last control flow instruction it depends on commits. Note that only WFC prevents Meltdown-style attacks which do not depend on a branch misspeculation, but it is possible that other defenses can cover Meltdown since only Intel processors appear to be vulnerable to it. *SafeSpec* makes no assumptions on the branch predictor behavior or on speculative execution behavior; for example, it does not prevent the attackers from mis-training or even polluting the branch predictor, nor does it prevent them from speculatively reading privileged data. Rather, SafeSpec interferes with the attacker's ability to create a covert channel using speculative data accesses to communicate illegally-accessed data out. We describe *SafeSpec* in Section III.

We demonstrate the *SafeSpec* principle by building a memory hierarchy (caches and TLBs) that are free from speculation-induced leakage. Making the memory hierarchy speculation-leakage free prevents Meltdown and Spectre attacks. In particular, we expand the load-store queues to store a pointer to a temporary associative structure that holds speculatively loaded cache lines. We also introduce a similar structure to hold speculatively loaded TLB entries. We describe the design and some of the complexity-performance trade-offs in Section IV.

Although the *SafeSpec* design presented so far completely prevents direct leakage of speculative side-effects, we identify a transient type of leakage that occurs in the introduced speculative state. Specifically, since many instructions start out as speculative before they commit, there is a period of time that eventually committed instructions share the shadow state with eventually squashed instructions. We show that a covert channel can be created in this transient period. We call this type of attacks, which is a byproduct of *SafeSpec*, *transient speculation attacks* (TSAs). We explore how to construct the shadow state to mitigate TSAs in Section V. The resulting design prevents all possible speculative leakage in the CPU caches and TLBs.

Section VI presents a performance, complexity and security analysis of *SafeSpec*. We show that both WFC and WFB models result in negligible impact on performance (in fact, modest improvements for some benchmarks). Although we miss out on some prefetching benefits of misspeculation [11], we benefit from the extra structures that hold speculative data, effectively increasing the size of the cache and the TLBs, and the reduction of pollution of the primary structures. We also analyze the complexity of *SafeSpec* including the impact of all new structures, and demonstrate a reasonable increase in the area and power consumption. Finally, we show that *SafeSpec* stops proof-of-concept implementations of all three variants of Meltdown and Spectre, as well as the new variants that we introduced.

Designing speculative architectures that are leakage free in a principled way requires carefully rethinking most aspects of the processor microarchitecture. This paper takes a first step into exploring this problem, and by necessity leaves many steps to future research. Specifically, we show *SafeSpec* only for the memory translation and access components of the CPU, which closes most currently known attack variations. To protect the processor and the full system, all speculatively updated structures, such as the branch predictor and the DRAM buffers must be protected [5], [12], [13]. Having established the *SafeSpec* principles, future work should carefully identify how speculative instructions can affect the state of the system and apply the same principles to protect them. The default simulation models in the simulator we used (MarSSx86) do not use a prefetcher. In the case when a prefetcher is used, we should include all the lines brought in by the prefetcher as part of the speculative state. Protecting the prefetcher completely requires delaying its training so that its influenced only by memory accesses that end up committing. Another limitation of our work is that we consider only single-threaded workloads. Multi-threaded workloads introduce issues of cache coherence and memory consistency model, which are solvable. However, they also introduce their own versions of speculative attacks [14] that must also be handled. Thus, we elected to focus this paper on the single-threaded model to establish feasibility and to gain initial experiences. We discuss these and other limitations of our solution in Section VII.

In summary, the paper makes the following contributions:

- We introduce the *SafeSpec* model to protect against speculation attacks by separating speculative state into separate structures.
- We design a leakage-free cache hierarchy using this principle that mitigates all three known variants of the attack. We identify a number of performance and complexity issues and introduce solutions for them.
- We identify a new class of speculative attacks (Transient Speculation Attacks) that arises in *SafeSpec* implementations where the covert exchange of information occurs completely while the instructions are speculative. We mitigate this class of attacks by sizing the shadow structures to prevent contention.
- We evaluate *SafeSpec* for caches and TLBs from a performance and complexity perspective. We show that the hardware overhead is acceptable and that we are able to retain the performance benefits of speculation (in fact, performance is slightly improved due to the additional cache space in the shadow state).

## II. Speculation Attacks and Threat Model

In this section, we introduce speculation based side-channel attacks (speculation attacks for short). The section first discusses speculative execution to characterize the capabilities of the attacker, and then overviews the Meltdown and Spectre attacks.

### A. Speculative Execution in Modern Processors

Speculative execution has been an important part of computer architecture since the 1950s. The IBM Stretch processor implemented a predict not-taken branch predictor [15]. As computer architecture continued to advance rapidly, the amount of speculation that is exploited has progressively been increasing with aggressive out-of-order execution, supported by sophisticated branch predictor designs [16], [17], [18] that are highly successful in predicting both the branch direction and its target address. In particular, the number of pipeline stages in production CPUs has continued to grow to the point where modern pipelines commonly have between 15 and 25 stages. Moreover, with out-of-order execution, when a branch instruction stalls (e.g., due to a cache miss on which it depends), instructions that follow the branch are continuously being issued. Thus, the speculation window can be extremely large, typically limited by the size of structures such as the reorder buffer, which can hold a few hundred instructions.

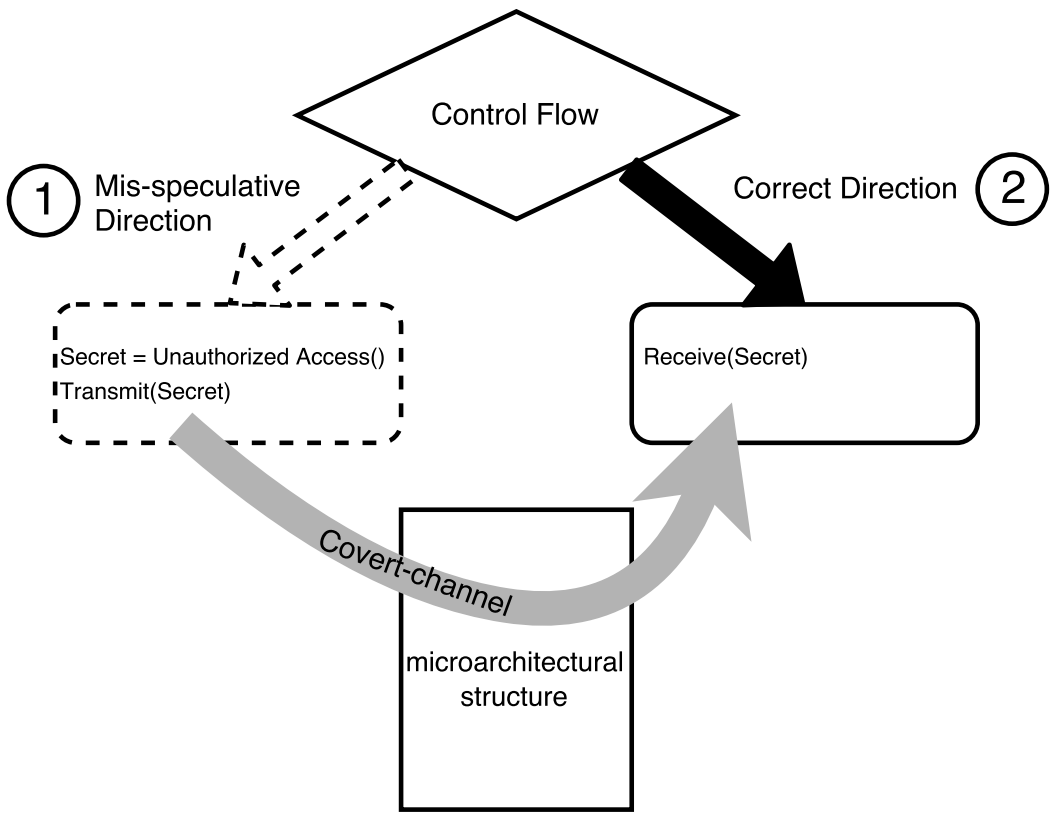

Fig. 1. Attack 1

Speculation is designed to not affect the correctness of a program. Although branch mispredictions occur and speculative instructions can ignore execution faults (e.g., permission error for memory access) these semantics were not considered harmful as mis-speculation will eventually be detected and the uncommitted instructions will be squashed, leaving no directly visible modifications to the architectural state such as registers and memory. Micro-architectural structures such as caches and TLBs are affected by speculative operations, but the contents of the cache only affect performance, not the correctness of a program. In fact, prior work has shown that there are beneficial prefetching side-effects to speculatively executed instructions even those that are eventually squashed [19]. To exploit these effects, designs such as runahead execution [11] intentionally increase the speculation window beyond the physical limitations of the reorder buffer to generate additional cache misses further into the program to exploit their prefetching effect. This approach was shown to significantly improve single-threaded performance.

### B. *Speculation Attacks*

Meltdown and Spectre are two representative attacks of the class of speculation attack. In general, these attacks exploit three properties of speculative execution in modern processors:

- **P1**: branch prediction validation and permission checks are performed deep in the pipeline and execution fault is generated only if the instruction is committed, enabling speculative instructions to access data outside its privilege domain;
- **P2**: speculative instructions leave side-effects in micro-architectural structures such as caches, which can be inferred using well-known techniques like Flush+Reload [20] and Prime+Probe side-channel attacks [21].
- **P3**: the branch predictor can be mistrained (Spectre 1), or directly polluted (Spectre 2). It is shared across all programs running on the same physical core [2], [3], [5], allowing code running in one privilege domain to manipulate branch prediction in another domain (e.g., kernel, VM, hypervisor, another process, or SGX enclave).

Next, we overview how different variants of Spectre and Meltdown attacks work, and distinguish them based on how they trigger and leverage speculative execution.

```
unsigned char secret;
dummy  = array[secret * 64];
```

Fig. 2. Secret-revealing gadget.

*1) A Common Gadget:* Speculation attacks aim to "read" memory/register content that is otherwise restricted. Unlike traditional memory reads, speculative reads are based on triggering speculative execution of a small code chunk, called gadget. A simplified example of such gadget is demonstrated in Figure 2. Assume the variable `secret` holds a secret value and is used as an index into a byte `array`. If processor speculatively executes this code, a memory access will be generated, and as a result some data will be borough into the data cache. Note that the secret variable controls what cache set will be updated by the speculative execution hardware. The multiplication operation ensures that different values of the variable will result in different cache sets. Because in many cases the location of data and code structures in victim process memory is not secret, the attacker, capable of monitoring CPU cache activity can link the observed behaviour with the corresponding value of the secret variable. For example, by knowing cache set $s_0$ is accessed when the value of `secret` is 0, the attacker can deduce that the value of `secret` must be $1 (\bmod n)$, with $n$ equals to the number of sets in the cache, when an access to the cache set $s_1$ is detected. Cache updates can be detected by attacker using a range of cache side channel attacks [20], [21], [22]. Please note that in normal execution, this code will never be executed, otherwise it will result in trivial cache side channel leakage. In speculative attacks the attacker uses the properties **P1** and **P3** described above to trigger the gadget to be *speculatively* executed by the victim.

```
if (offset < array1_size)
    y = array2[array1[offset] * 64];
```

*2) Spectre (Variant 1):* This variant of the attack can be demonstrated by the code presented in Figure II-B1. In this code, a victim process reads values from `array1` using the `offset` provided by the attacker. Then, resulting value is used to perform an access into `array2`. As we discussed above, accesses into the `array2` can be used by the attacker to deduce the value of the index. The index, in its turn, is controlled by the attacker since attacker controls the `offset`. Therefore, the attacker can use a carefully selected value of `offset` to read arbitrary memory address which then will result in cache access observable by the attacker. However, the `if` statement ensures there are no out of bounds memory accesses allowed. Unfortunately, the attacker can exploit speculative execution and behavior of branch predictor to force the victim process to perform an out of bounds memory access in the following way:

a) The attacker triggers the code to be executed several times and with the value of the `offset` such that the `if` statement is always true *(branch instruction not-taken)*. This trains the branch predictor to predict the corresponding branch always not-taken;
b) Next, the attacker flushes `array1_size` from the cache, forcing the CPU to fetch the value of `array1_size` from memory, delaying the correct evaluation of the branch and creating a large speculative window;
c) Finally, the attacker provides the malicious `offset`. The branch predictor unit predicts the branch not-taken, resulting in two memory accesses that reveal the value stored at the attacker's desired address.

*3) Spectre (Variant 2):* In this variant of the attack, the target program may not contain the expected gadget or the `offset` is not controllable by the attacker. These limitations can be bypassed by *hijacking* the speculative execution. Specifically, when the CPU encounters an indirect branch instruction, the branch predictor tries to guess the target address and the CPU immediately starts speculatively executing instructions at this address. Due to **P3**, the attacker can perform the *branch target poisoning* to hijack the speculative execution flow and to redirect it to any code location containing gadget instructions. This resembles the return-oriented programming attack [23]. In summary, the variant 2 attack works as follows:

a) The attacker ensures that the attacking code and the victim code share the same branch target buffer (BTB) by executing attacker's process on the same physical core with victim.
b) The attacker forces a BTB collision by matching virtual address of the victim and attacker branch instructions [5].
c) The attacker performs target poisoning by executing its own branch.
d) Finally, the attacker triggers the indirect branch to be speculatively executed, redirecting the speculative execution to a gadgets of attacker's choice. The gadget will leak data through a side channel in a way similar to previously described.

*4) Meltdown:* Meltdown attack exploits **P1**: due to pipelining and instruction reordering a permission check can happen *after* the corresponding memory accesses is speculatively executed. For example, assume a user application that tries to read kernel memory. Although such request will be eventually denied, the speculatively executed instructions will result in loading of requested data into caches. Using a side channel, the attacker can effectively read arbitrary kernel (or hypervisor) memory. This is a very powerful attack, since typically kernel memory contains a direct mapped region allowing the attacker to dump the entire physical memory on a given system. Since the exception eventually will be raised, this attack requires the ability to tolerate and recover from segmentation faults. Alternatively, if the attacker can arbitrarily control the exploit code, she can also avoid the exception by putting the gadget behind a mispredicted branch, i.e., combining Spectre V1 with Meltdown to read memory across privilege domains in the same virtual address space.

### C. Threat Model

Since **P2** is essential in all speculation attacks, this work aims to eliminate the side-effects from speculative execution. Hence, we assume a strong adversary for the branch predictor and no software-based defense for branches (e.g., `lfence` and retpoline [7]). In particular, we assume that attackers can arbitrarily control the state of the branch predictor, as if its state is programmable without any privilege. We assume attackers can launch a speculation attack either from the same process or another process. We assume attackers have complete control over the attacking code (as in the Meltdown attack) and know the complete layout of the victim domain (another process, kernel, enclave, etc.). Their goal is to reveal memory and/or register content of the victim domain. We assume the victim domain does not have any direct channel or vulnerability to leak the content so attackers must utilize side-channels. To enable these side-channel, we assume the victim domain contains the code gadget such as the one in Figure 2 that can be invoked by attackers.

The technical solution we propose is general and applicable to different micro-architectural structures. However, as a demonstration, our prototype implementation only protects caches and TLBs to explore concretely the implications and complications that result from *SafeSpec*. Therefore, we further assume that other covert channels, including the ones through the branch predictor, memory bus and DRAM buffers are out-of-scope for the current paper, but will be addressed using similar principles by future work. Similarly, we only consider a system with a single core. Thus, speculation attacks against the cache coherence and memory consistency model states [24] are also left for future work. We discuss the implications on both of these in Section VII.

## III. SafeSpec: Leakage-free Speculation

This paper proposes a principled way to secure processors against speculation attacks while retaining the ability to carry out speculative execution to benefit from its performance. The general principle (shown in Figure 3) uses temporary structures (shadow state on the figure) to hold any state that is produced speculatively without affecting the primary structures of the processor (which we call committed state on the figure). For example, if a speculative load instruction causes a load of a cache line, instead of loading that cache line into the processor caches, we hold the line in a temporary structure. If the load instruction is later squashed, these effects are removed in place (bottom path from the commit state), leaving no changes to the cache from the mis-speculated instructions, and closing the vulnerability. Alternatively, if the instruction commits, the cache line is moved from the temporary structure to the L1 cache and removed from the shadow state.

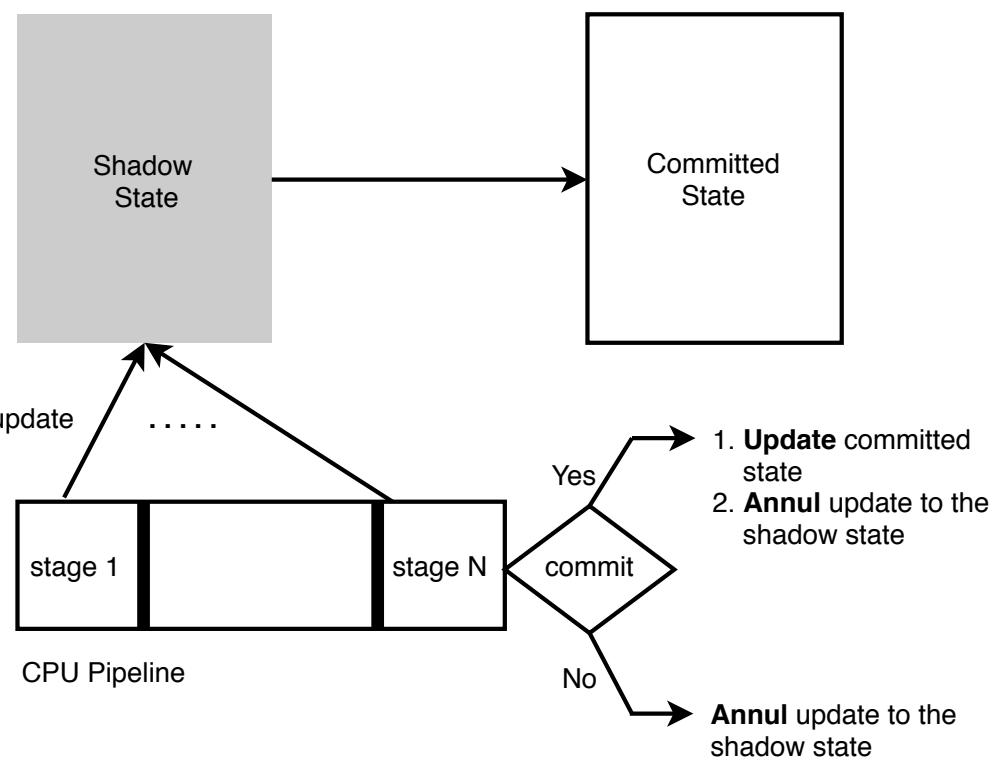

Fig. 3. SafeSpec overview

While *SafeSpec* is simple in principle, a number of questions relating to its security, complexity and performance have to be resolved. We overview these issues in the remainder of this section.

**When to move state from speculative to committed.** There are two options available to decide when to move state from the shadow to the committed state. In the first variation, which we call *wait-for-branch* (WFB), we can assume an instruction to be no longer speculative when all the branches (more generally, all predictions) it is dependent on have been resolved. WFB stops both variants of spectre which depend on mistraining the branch predictor; none of the mis-speculated instructions moves to the committed state. However, it does not prevent Meltdown which does not rely on the branch predictor.The second variation *wait-for-commit* (WFC) waits until the instruction causing a speculative side effect commits before moving its effects to the committed state, and therefore also prevents Meltdown. We note that the Reorder buffer according to this definition is a form of shadow state, whose data is moved to the permanent state (architectural state) only when the instruction is committed.

**Shadow state organization and size:** If the shadow state structures are too small, then either speculative state is replaced (causing a loss of an update to the committed state if this data were to be committed later), or the instruction has to stall until there is room in the speculative structure before it issues. Thus, from a performance perspective, the organization and size of the shadow structure should be designed such that the structures can hold the speculative state generated by speculation as measured across typical workloads. However, we will show that security considerations introduce more stringent requirements on the speculative state.

**Mitigating Transient Speculation Attacks:** *SafeSpec* by construction prevents speculative values from affecting the state of committed structures, which is the pathway used to communicate data covertly in the published speculation attacks. However, it does not create isolation between instructions that are in the speculative state. This creates a possibility for a new variant of attacks which we call *transient speculation attacks (TSAs)*. In particular, since instructions that commit can be in the speculative state (before their dependent branch commits in WFB, or before the instruction itself commits in WFC), there is a window of time where they can share the speculative state with misspeculated instructions before they are squashed. If we are not careful, it is possible to create a covert channel in this period to communicate the sensitive data from the mis-speculated branch to the branch that will be committed, allowing the data to be exfiltrated.

Consider the example of a shadow structure that is sized to be small (lets say one entry). The malicious speculative code that reads the privileged data can then communicate it covertly to speculative code (the ”receiver” code that will commit) using the shadow state. For example, it can replace the entry in the shadow state, causing the receiver to notice the absence of its speculative state (since it was replaced) after it commits. Alternatively, if we block when the shadow structure is full, the receiver can detect that its code took a longer time to execute.

Although TSAs are strictly less powerful than the original attack, they must be carefully considered to ensure that leakage is not possible. One way to solve this problem is to either partition the speculative state per branch, or to size it generously, or even for the worst case scenario, to ensure that no leakage occurs through the shadow state. TSAs can also attempt to communicate covertly by creating contention on functional units or other shared structures; this is an issue that we also consider. We discuss how to mitigate TSA attacks in Section V.

**Filtering Delayed Side Effects:** One of the issues with *SafeSpec* occurs when an instruction is squashed in the middle of its execution. If the instruction has already initiated a high latency operation such as a read from memory, we have to ensure that the response from memory can be discarded after it was received. The instructions executing speculatively store any resulting state to the shadow structures. Thus, if a long latency reply is received and there is no matching transaction, we simply discard these values. However, it may also be desirable to filter these transactions lower in the system, such that the committed transactions commit directly, and the squashed ones are cancelled in place. To control the size of this filter, we can include a branch id with the transactions and track operations at the branch granularity. The filter can also be used to mark committed branches so that memory responses corresponding to them are committed directly to the permanent structures.

## IV. SafeSpec for Caches and TLBs

To demonstrate the *SafeSpec* principle, we implemented it to protect CPU caches and TLBs from leakage during speculative execution. We chose the CPU caches because they are easily exploitable targets for covert communication and the ones used in the Spectre/Meltdown attacks. Caches have simple indexing and with the availability of instructions such as `clflush` on x86, an attacker is able to evict data which facilitates

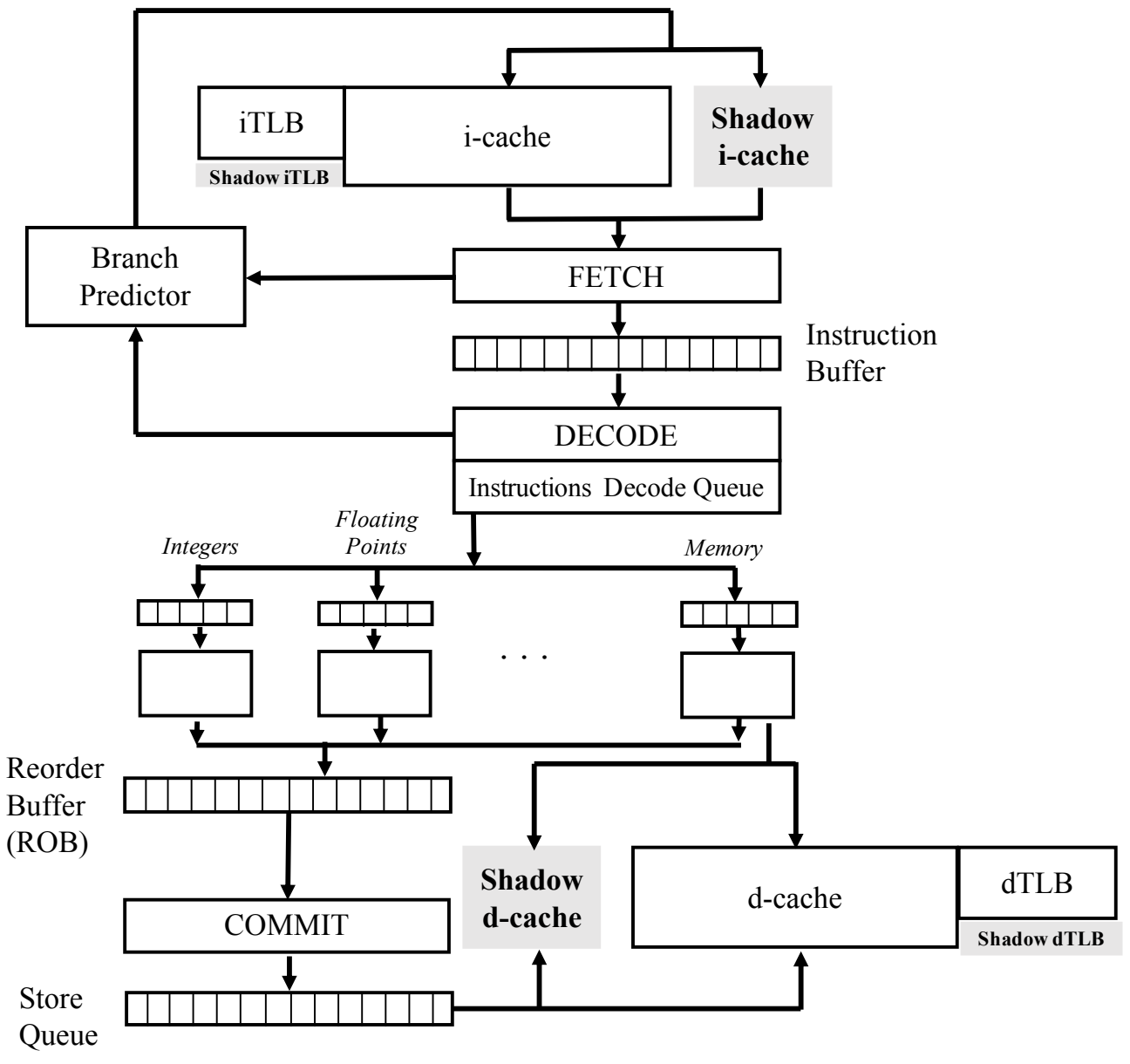

Fig. 4. Safespec Extension to the CPU Pipeline

quick exfiltration using, for example, a flush+reload covert channels [20].

### A. Shadow Structures

To protect from speculative covert channels that occur during memory accesses, and following the *SafeSpec* principles, we need to add shadow state to protect the following structures (Figure 4).

- Data caches: this is the covert channel used in all three Meltdown/Spectre variants. We add a shadow structure to hold the cache lines that have been fetched speculatively. The structure is filled associatively, but accessed as a lookup-table. In the Load/Store queue, we point speculative loads that have received their data to a corresponding entry in this table. Speculative instructions in the *same execution branch as the load that fetched a shadow cache line* that access this cache line can use the value from the shadow structure. If an instruction commits (depending on WFB or WFC), the cache line is moved from the shadow structure to the caches. If the instruction is squashed, the shadow structure entry is marked as available. Note that this way, not even the cache replacement algorithm state is affected by the speculative data that does not commit.
- Instruction cache: we built a variant of Meltdown/spectre using the instruction cache. This variant, conceptually shown in Figure 5 replaces the data dependent array access with dependent branches (in function gadgetFunc) to a location in an array to disclose the data through the I-cache. To develop this attack variant, we had to overcome several challenges. In particular, the data dependent branches use the branch predictor, but the I-cache footprint from this branch is not data dependent because the value in the BTB is not data dependent either. Thus, we had to initialize the BTB to a third location, and then introduce sufficient delay in the pipeline for the data dependent branch to be resolved such that it registers the data dependent location in the I-cache. Our working attack on the I-cache demonstrates the need to protect the I-cache, which can be accomplished similar to the D-cache.
- TLBs: we also conjectured that the TLBs may be used as a covert channel vector. Essentially, the data dependent access would target a page based on the value of the data, causing the corresponding TLB entry to be initialized. Later, we can check the time to access the page to see if it results in a TLB miss or not to derive the communicated data. In fact, the translation pathway in modern CPUs is quite complicated, as recent reverse-engineering effort demonstrated [25]. The memory datapath includes including additional translation caches, the page directory entries and page table entries are also brought into the data cache. Fortunately, the page walker uses the load-store queue for these accesses, and the protection introduced for the data caches ends up protecting these structures as well, thus the additional protection is only needed for the TLB itself.

```
attackMode <-- 0;
secret <-- readSecret();
int (*fnPtr)[256 * 256];
for all ascii in (ASCII-character)
{
    define int func_ascii() //noop sled
    {
        asm volatile(".rept 256;" "nop;"".endr;");
        return 0;
    }
    (*fnPtr)[ascii * 256] <-- func_ascii
}
clflush(&array1_size);
clflush(fnptr);
function speculative(secret)
{
    if (secret < array1_size)
        gadgetFunc(secret, attackMode);
}
function gadgetFunc(secret, attackMode)
{
    if (secret == 'A' && attackMode)
        fnPtr[A*256]();
    // ... 256 If Structures for
    // all ASCII characters
    if (secret == 'z' && attackMode)
        fnPtr[z * 256]();
junkLoc();

}
for (i = 1...256)
{
    t1 = rdtscp();
    junk = fnptr[i * 256](); //check cache hit
    t2 = rdtscp();
}
```

Fig. 5. New I-cache variant of Spectre

Thus, we have to add shadow structures to protect these three components of the memory system. In the remainder of this section, we discuss the implementation and the sizing of these structures from a performance perspective. We consider both commit policies: WFB and WFC.

### B. Implementing and Sizing Shadow Structures

To implement *SafeSpec* for the data cache, we add an associatively-filled lookup table to hold speculatively read cache lines. It is important to note that memory consistency models, such as Total Store Order (TSO) semantics of the x86-64, often ensure that store side-effects appear in order; in other words, the cache is not updated until the store commits, making stores robust to speculation attacks. We augment the load store queue with a pointer to the shadow cache line for load operations that are speculative. Any instruction dependent on the speculative load reads the cache line from the shadow structure. Once the load instruction commits, the shadow cache line is written to the caches according to the inclusion policy of the caches (in our case, since the caches are inclusive, it is written to all levels of the cache) and freed in the shadow structure. If the load is squashed, the value is freed in the shadow structure. For the i-cache and the TLBs, we create similar shadow structures, and augment the ROB with pointers to the shadow state entries if the instruction is speculative and the cache line (or TLB entry) were fetched speculatively.

From a performance perspective, the structures should be sized such that they accommodate the speculative state needed by representative workloads (e.g., measured empirically). If the shadow structures are full, we could either drop some of the shadow state (leading to loss of updates to the committed state with performance, rather than correctness implications), or block until there is space in the shadow state before issuing an instruction (also with performance implications). It is interesting to characterize the size of this state, but we will see later that the constraints introduced by security requirements to eliminate TSAs are more stringent than those required by performance. Figures 6 and 7 show the distribution of the size of the speculative state sampled over time for the SPEC 2017 benchmarks for a processor configuration similar to the Intel SkyLake processor. The shadow d-cache for 3 of our benchmarks grows occasionally to almost the maximum possible size (bound by the size of the load-store queue). A shadow I-cache with about 25 cache lines is sufficient for all of the benchmarks. Figures 8 and 9 show the speculative TLB state for the same benchmarks. Less than 10 entries are sufficient for speculative iTLB misses, but some benchmarks require more d-TLB entries (up to 25). Overall, it is interesting to observe that the benefit from doing WFB is small, so we elect to support WFC to get the increased protection to cover Meltdown. In the next section, we also show that to protect against the potential of transient attacks in the shadow state, we need to size the shadow state larger than what is required just for performance.

## V. Transient Speculation Attacks (TSAs): Covert Channels in the Speculative State

The *SafeSpec* principle prevents direct side-channel leakage from the speculative state to the committed state, closing all known speculation attacks. However, although the committed instructions and the speculative instructions eventually reside in separate structures, creating the separation and closing the channel, eventually committed instructions can start out as speculative. During this window, the eventually committed instructions share the shadow state with any speculative instructions that will be squashed. If the shadow structures are not designed carefully, covert channels can be created during this transient window to communicate sensitive data (which can only be read by a mis-speculated path) to an instruction pathway that will be committed such that the leakage results are visible to the program.

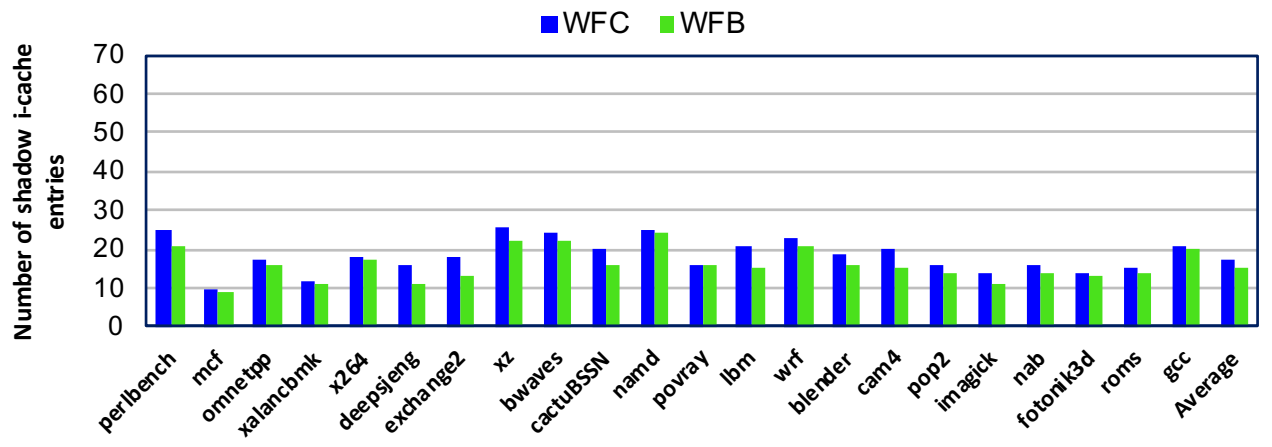

Fig. 6. Size of shadow i-cache that can fit 99.99% of the accesses

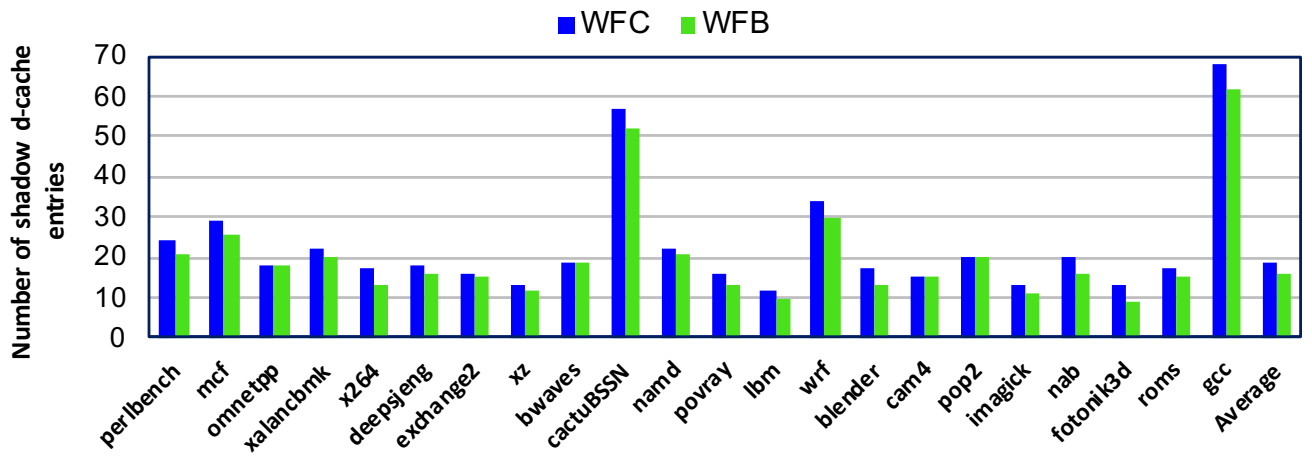

Fig. 7. Size of shadow d-cache that can fit 99.99% of the accesses

It is important to emphasize that these attacks (which we call Transient Speculation Attacks, or TSAs) are substantially more difficult than Spectre/Meltdown because there is only a limited window of speculation in which the malicious Trojan code must not only read sensitive data, but also create measurable contention to the spy before either of their predicate branches commits. In the Spectre attacks, malicious speculative code (the Trigger) is only triggered with the responsibility to delay the resolution of dependencies that the Trojan's branch relies on to allow the Trojan to read the privileged data and create a dependent access in the cache. This access is later checked by the spy code, at its own pace and outside the speculation window used by the Trojan. In contrast, in TSAs, the full attack has to happen within the speculation window during which both the Trigger and the Spy must be executing. In other words, the Trigger must also act as the Spy and receive the sensitive data from the Trojan inside the speculation window, making the attack substantially more difficult to perform.

The next question to consider is: what are the covert channels available for TSAs to communicate between the Trojan and the Spy? Since the primary processor structures are now separated from the speculative state, the attackers can try to use the shadow structures to communicate covertly. In particular, if the shadow structures are shared and sized such that they enable contention, they can be used for covert communication.

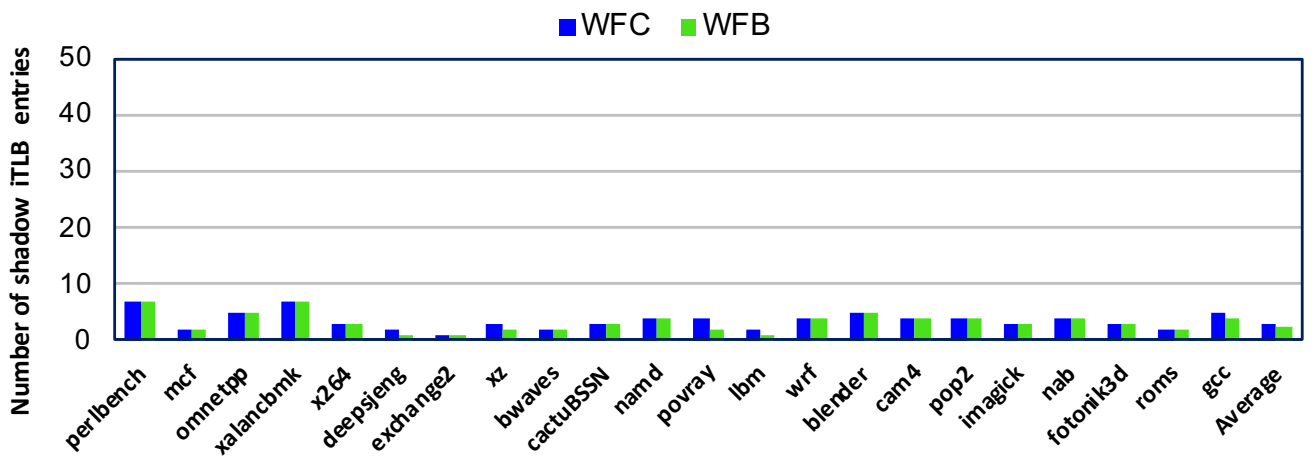

Fig. 8. Size of shadow iTLB that can fit 99.99% of the accesses

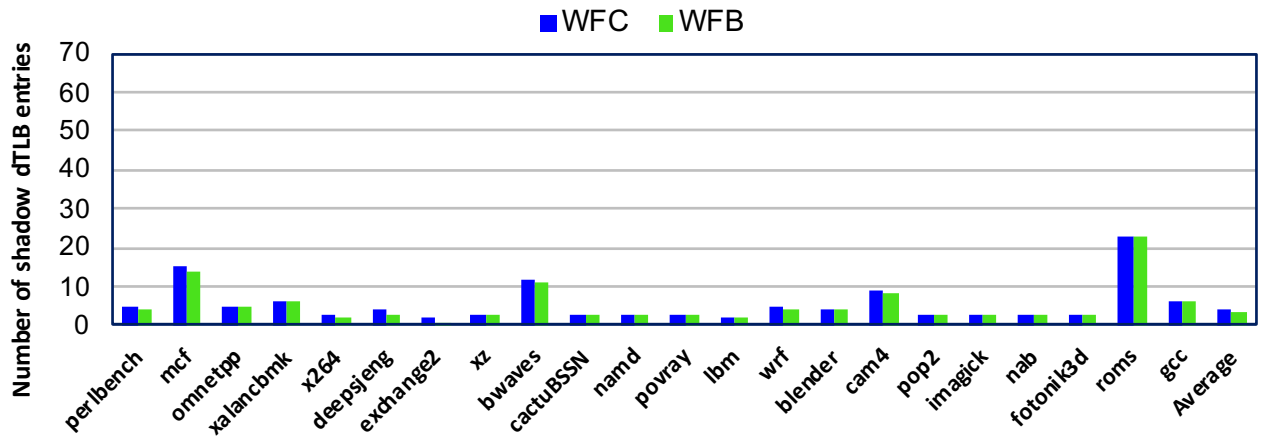

Fig. 9. Size of shadow dTLB that can fit 99.99% of the accesses

TABLE I
CONFIGURATION OF THE SIMULATED CPU

| **Parameter** | **Configuration** |
|---|---|
| CPU | SkyLake |
| Issue | 6-way issue |
| IQ | 96-entry Issue Queue |
| Commit | Up to 6 Micro-Ops/cycle |
| ROB | 224-entry Reorder Buffer |
| iTLB | 64-entry |
| dTLB | 64-entry |
| LDQ | 72-entry |
| STQ | 56-entry |

TABLE II
CONFIGURATION OF THE SIMULATED CPU MEMORY SYSTEM

| **Parameter** | **Configuration** |
|---|---|
| L1I-Cache | 32 KB, 8-way, 64B line, 4 cycle hit |
| L1D-Cache | 32 KB, 8-way, 64B line, 4 cycle hit |
| L2 Shared Cache | 256 KB, 4-way, 64B line, 12 cycle hit |
| L3 Shared Cache | 2 MB, 16-way, 64B line, 44 cycle hit |
| Memory | 12 GB, 191 cycles |

Consider an example where we size the TLB shadow structures based on typical program behavior. Since programs do not have many pending TLB misses within a speculation window, it stands to reason to size these structures to be small. In the rare case when the shadow structures are full, we may handle this by either discarding updates or by blocking the issue of requests when there is no room in the shadow structure. Either of these behaviors provides potential for a covert channel. Consider that the Trojan fills the structures with TLB misses if it wants to communicate a 1. If updates are discarded, a spy can detect a communication if its TLB accesses are not committed (they were discarded). Alternatively, if we block TLB accesses when the structures are full, the spy can detect a communication of 1 if its TLB accesses are delayed causing a longer TLB miss time. The attack is illustrated in Figure 10.

To prevent TSAs through the shadow structures, we can either partition the structures such that there is no contention among different speculative branches or we could provision them generously (even for the worst case) to make sure that transient contention cannot be created within a speculation window. As a proof of concept, we elect this second approach where we derive the maximum possible size of each shadow structure and provision for this worst case. This approach guarantees that no contention on the shadow structures is possible, at the cost of provisioning fairly large associative structures. We believe that with some more analysis, or with some detection defense that detects an attack when the shadow structures grow abnormally large, this worst case provisioning can be substantially relaxed without introducing leakage.

## VI. PERFORMANCE, SECURITY, AND OVERHEAD EVALUATION

To evaluate the proposed *SafeSpec* design, we conduct experiments with MARSSx86 (Micro Architectural and System Simulator for x86) [26]. MARSSx86 is a fast cycle-accurate full-system simulator that uses PTLsim [27] for CPU simulation on top of the QEMU [28] emulator. It is targeted to simulate cycle accurate out-of-order x86 cores. It allows users to capture separate statistics of user mode and kernel mode activities in a single simulation run. The CPU and cache model configurations of MARSSx86 were modified to simulate Intel Skylake processoras shown in Table I and Table II respectively.

### A. Performance Analysis

The first experiment measures the performance of *SafeSpec* compared to the baseline processor under conservative condition. In particular, we consider the shadow state access time to be equivalent to the access time of the L1 cache (4 cycles), when it is substantially smaller, and accessed as a lookup table. Figure 11, shows the IPC values for all SPEC2017 benchmarks. We see a small improvement in performance with a geometric mean of about 3%. We believe that this advantage results from a combination of effects including the larger effective cache size and avoiding polluting the cache with wrong path speculative state.

To gain more insight into the observed performance, Figure 12 shows the miss rate on read operations in the D-cache. There is little difference in behavior between SafeSpec and the baseline with respect to the data accesses. Figure 13 shows the percentage of the reads that hit the shadow structures.

The I-cache behavior is significantly different than the D-cache. Figure 12, and 14 shows the miss rate on the I-Cache. For the I-cache, there are more substantial differences between WFC and the baseline. Some outlier behavior such as Pop2 and imagick where the percentage of i-cache misses drops significantly could be due to the larger size of the shadow structures expanding the effective size of the cache reducing conflict and capacity misses. Moreover, we see in Figure 15 that most of the hits occur in the shadow i-cache structure reflecting the high spatial locality of the access patterns in the i-cache; in other words, while a cache line is still speculative, several instructions execute from the same cache line. In contrast, the D-cache has less spatial locality, resulting in fewer accesses hitting the shadow state. We note that the cache miss rates are combined for all instructions (i.e., we do not exclude instructions that are squashed); therefore, many of these hits in the shadow structures may not end up being productive.

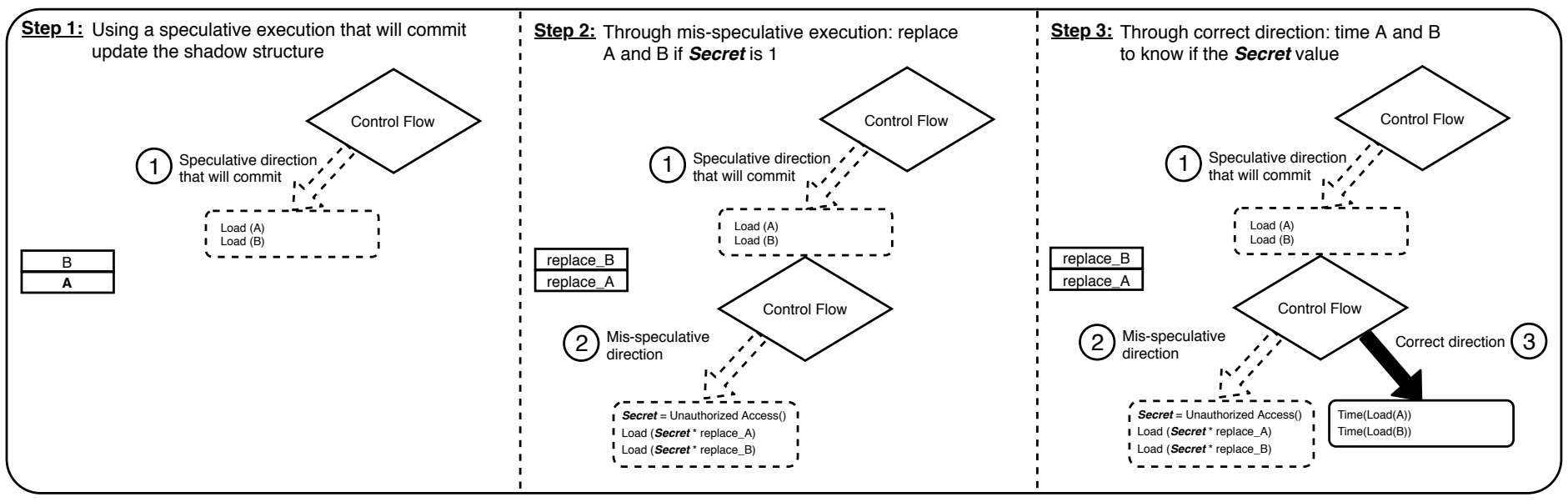

Fig. 10. Transient Covert Channel Attacks

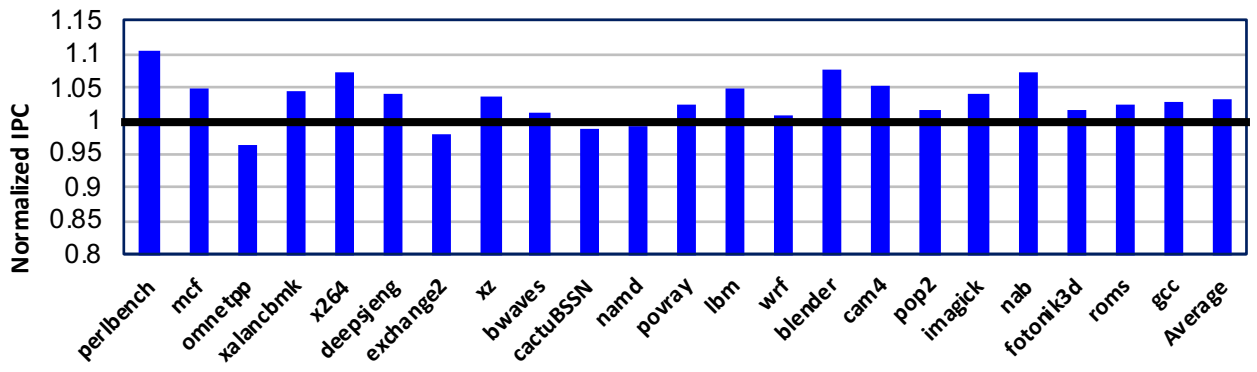

Fig. 11. Relative Performance to Non-secure OoO Execution

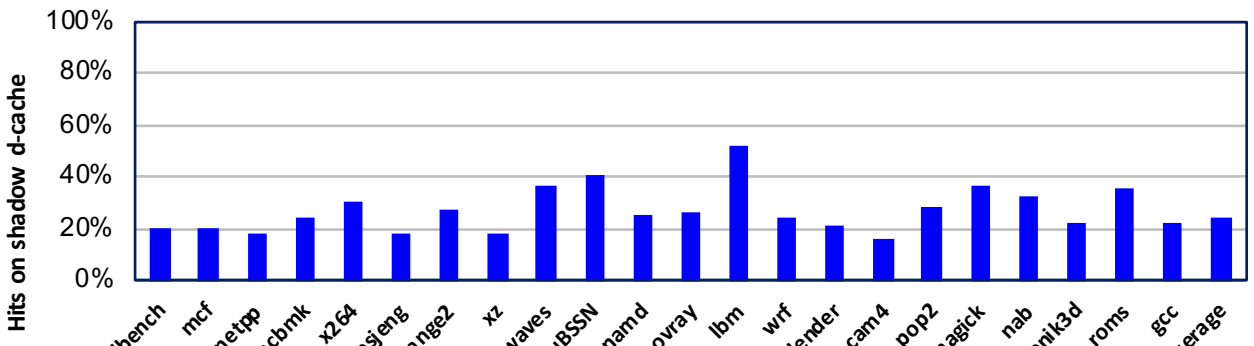

Fig. 13. Percentage of hits on shadow d-cache

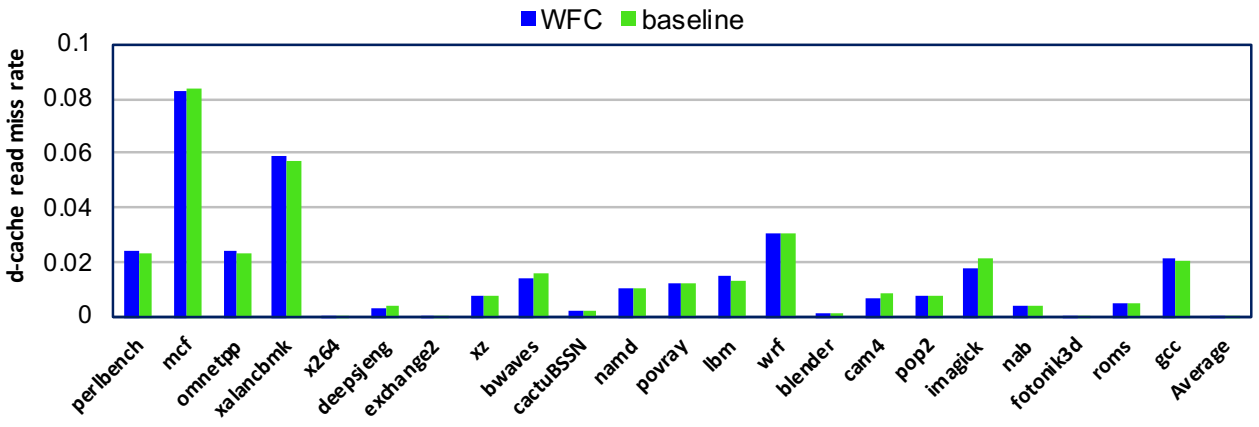

Fig. 12. d-cache read miss rates including the shadow d-cache

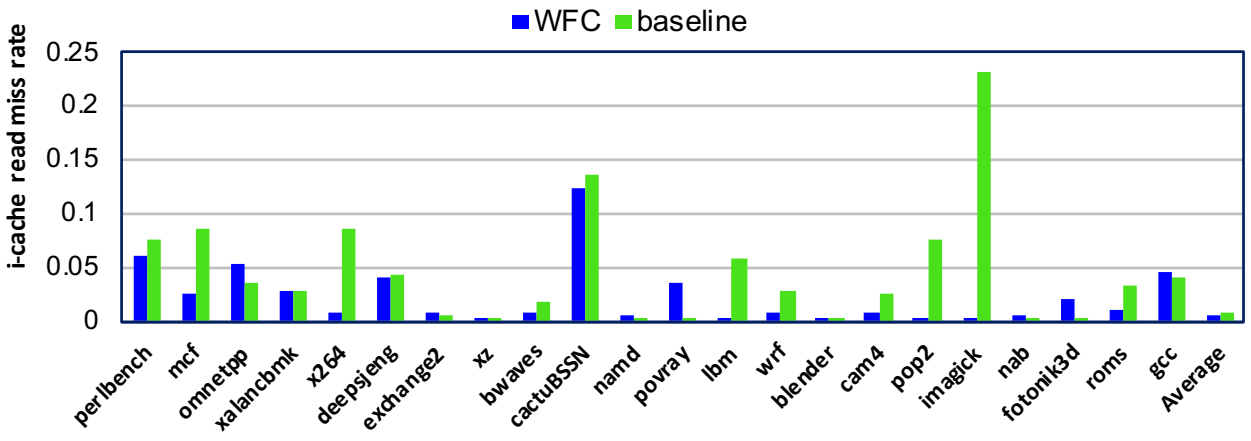

Fig. 14. i-cache miss rate including the shadow i-cache

To understand the benefits of the shadow structure in filtering misspeculated accesses, Figure 16 shows the percentage of the shadow state that ends up being committed for the I-cache and the D-cache. We observe that a substantially higher percentage of the D-cache state ends up being committed, perhaps due to the fact that speculative loads are issued later in the pipeline making them more likely to commit. For both the D-cache and especially the I-cache, the shadow structure filters a large number of misspeculated accesses that are squashed without cluttering the caches.

## *B. Security Analysis*

Table III shows that both WFC and WFB close Spectre attacks, but only WFC is guaranteed to also stop Meltdown attacks. We evaluated our proof of concept code implementing Spectre in the simulator and found indeed that the attack fails under both WFC and WFB models. Table IV shows the protection coverage for Spectre-style attacks targeting structures other than the d-cache (I-cache, I-TLB, and D-TLB). All three side channels were closed. We tested proof of concept code for the I-cache and a transient attack through the D-cache and observed that the attack fails on the *SafeSpec* protected CPU.

TABLE III
SECURITY ANALYSIS OF MELTDOWN/SPECTRE

| | WFC | WFB |
|---|---|---|
| Meltdown | ✓ | ✗ |
| Spectre 1/2 | ✓ | ✓ |

TABLE IV
COVERAGE OF SPECTRE STYLE ATTACKS ON DIFFERENT STRUCTURES

| | WFC | WFB |
|---|---|---|
| I-cache | ✓ | ✓ |
| I-TLB | ✓ | ✓ |
| D-TLB | ✓ | ✓ |
| Transient | ✓ | ✓ |

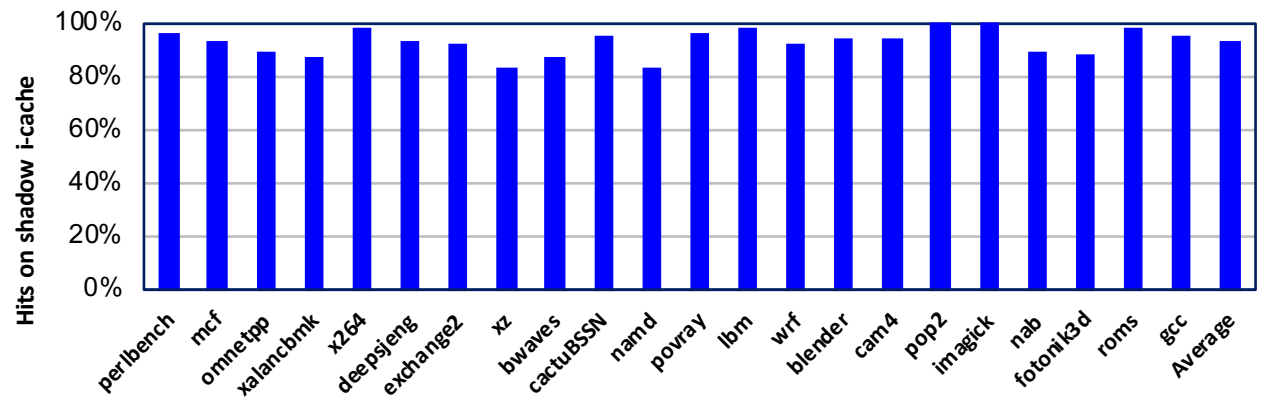

Fig. 15. Percentage of hits on shadow i-cache

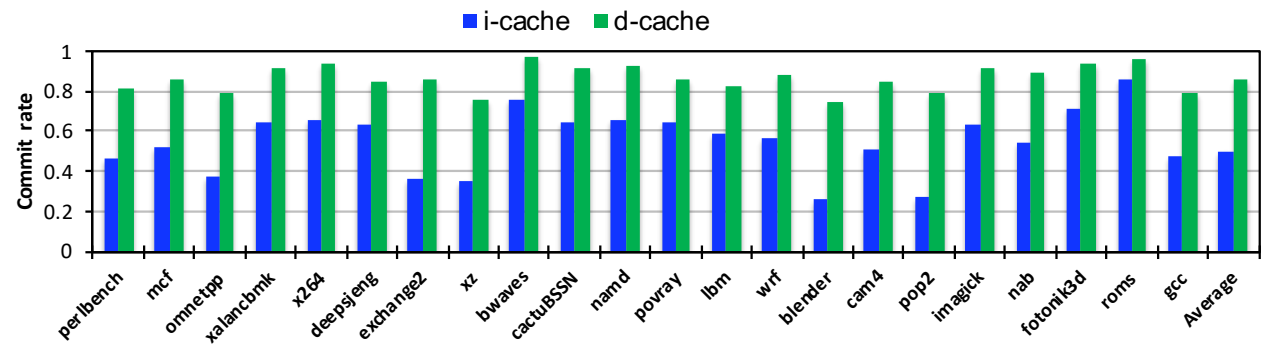

Fig. 16. Commit rate of shadow state

We could not get TLB-based attacks working in the simulator, perhaps because of the large delays of page walks, or due to the limitations of the MarSSx86 models of the TLBs.

### C. Hardware overhead

*SafeSpec* introduces hardware overheads to the CPU pipeline due to the addition of the shadow structures. We compared the hardware overhead for two different sizes for the shadow structures; 1) Secure: shadow structure size equal to the maximum speculative state during speculation; and 2) SafeSpec with WFC: shadow structure sizes were optimized based on 99.99% speculative state size for SPEC2017 benchmarks using the WFC implementation. We report the area, power, and access time values, as well as a percentage compared to the Skylake CPU L1 cache configuration (shown in Table II), using CACTI v5.3 [29] in Table V. The results show that the area overhead is tolerable for the secure design, making the design highly practical.

TABLE V
SafeSpec hardware overhead at 40nm.

| | Power ($mW$) | Power (%) | Area ($mm^2$) | Area (%) |
|---|---|---|---|---|
| Secure | 290.27 | 26.4 | 9.79 | 17 |
| WFC | 35.14 | 3 | 1.17 | 2 |

## VII. Discussion, Limitations and Future Directions

*SafeSpec* is a principled approach for protecting systems from speculation attacks by preventing crossover leakage from speculative instructions that will eventually be squashed to permanent structures where they could be visible to attackers through a side channel. By preventing this leakage, we close the covert channel that is exploited by recent speculation attacks such as Meltdown and Spectre. This general principle should be applied to all speculatively modified state within a CPU.

*SafeSpec* requires a deep redesign of the CPU to separate out the speculative state from the permanent state. It also has implications on security: we identified a form of transient side channels that occur through the shadow structures. The goal of this paper is to establish the *SafeSpec* principle by protecting the CPU caches and TLBs. We recognize that other structures affected by speculative instructions must also be protected using this principle or otherwise the attackers will switch to using them. Future work should look at protecting the branch predictor, DRAM buffers, account for prefetchers, as well as other structures.

Another limitation of the current work is that we do not support multi-threaded workloads. Addressing this limitation involves two considerations. The more straightforward consideration is how to preserve the semantics of protocols such as cache coherence, memory consistency models, atomic operations, and transactional memory. We believe that these continue to operate in the same way by treating the speculative state to be part of the state of the caches. The second issue is significantly more difficult: these protocols themselves can be used to communicate speculative side-effects as has been recently shown by the MeltdownPrime attack [24]. Designing leakage-free protocols is a difficult problem that deserves separate and complete treatment and therefore we elected to leave supporting multi-threaded workloads to future work.

We identified the problem of transient covert channels that occur while instructions that will eventually commit share the shadow structures with speculative instructions that will not. In the window while both set of instructions are speculative, they share the shadow state creating the potential for covert communication. To prevent covert communication, one approach is to size the shadow structures for the worst case contention level and make them fully associative. This worst case size is bound by the size of the load-store queue for the d-cache and d-tlb, or the size of the reorder buffer for the i-cache and the i-TLBs. While this pessimistic approach guarantees no potential for leakage, more careful analysis can show that a much smaller size will suffice given the transient nature of the exposure.

We also characterized the size of the shadow state created by normal program execution and showed that it is substantially smaller than the worst case. Thus, we expected these large shadow structures to be mostly unused providing opportunities for dynamically resizing them for energy efficiency. In addition, it is possible to use abnormal growth of the structures as an indicator of a possible attack and introduce mitigations to stop the attacks. This can also be explored in future work.

Speculation attacks challenge the foundation of out-of-order microarchitectures which have been the key building blocks of computer systems in the last several decades. Since these attacks are very new and most of the proposed defenses are at the software/firmware levels, CPU manufacturers and microarchitecture researchers face an open challenge of how to redesign speculative out-of-order processors to be immune to speculation attacks. This paper represents a first step in this direction that we hope will spur future research in this area.

## VIII. Related work

After the disclosure of Spectre and Meltdown in January, 2018 [1], [2], [30], a number of defenses were suggested.

**Intel proposed defenses:** Intel released a whitepaper [31] suggesting three types of defenses.

- To mitigate Spectre V1 attack, Intel recommends inserting a `LFENCE` instruction after the branch as a barrier to stop speculative execution. This defense mechanism has now been adopted by compilers such as GCC [32] and MSVC [33].
- To mitigate Spectre V2 attack, Intel introduced three new processor interfaces through microcode updates [34]:
  - Indirect Branch Restricted Speculation (**IBRS**) prevents software running in less privileged mode from impacting the indirect branch prediction of software running in more privileged mode (e.g., kernel and SGX enclave) or on another logical processor.
  - Single Thread Indirect Branch Predictors (**STIBP**) prevents software running on one logical processor from impacting the indirect branch prediction of software running on another logical processor or later on the same logical processor of the same core.
  - Indirect Branch Predictor Barrier (**IBPB**) establishes a barrier that prevents software running before the barrier to affect the indirect branch prediction of software running after the barrier.
- To mitigate Meltdown attack, Intel recommends unmapping more privileged domain (kernel space) during the execution of less privileged software, which has been adopted by all popular operating systems, including Windows, Linux, and macOS.

All of these mitigation mechanisms require software modifications or microcode updates, thus they can leave legacy systems unprotected. More importantly, these protections introduce non-negligible performance overhead which can range from 0–2% [35] to 800% [10] depending from application and hardware platform.

**Kernel Page-Table Isolation (KPTI)**: Gruss et. al [8] introduced a protection technique called KAISER to protect against side channel attacks bypassing kernel level address space randomization (KASLR) [36]. The protection is based on complete separation of kernel and user page tables. As a result, kernel and user programs exist in separate address spaces, effectively mitigating side channel attacks. It has been demonstrated that KAISER can effectively mitigate Meltdown [1] which is stopped because user applications no longer can perform speculative memory accesses to kernel address space since the kernel is completely unmapped. Such protection comes with a tangible performance overhead for system calls and context switches, due to frequent TLB flushes. As reported, the performance overhead can be as high as 800% [10] for system call intensive tasks, especially when running on older processors without support for process-based TLB tagging and selective invalidation. KAISER (also known as KPTI) was included in the mainstream Linux kernel as a response to the Meltdown attack [1]. The mitigation technique cannot protect from any variation of the Spectre attack or when Meltdown is performed within the same address space, for example in case of software modules protected with software fault isolation techniques [37], [38].

**Return Trampoline (retpoline)**: retpoline [7] is a software-based mitigation technique against indirect branch target injection attack (i.e., Spectre V2). It "exploits" two properties of the branch target prediction engine: (1) when executing a `ret` instruction, the predictor will utilize the return stack buffer (RSB) instead of the BTB; and (2) RSB cannot be polluted by attackers. The retpoline technique essentially swaps indirect branches for returns and deliberately pollutes the RSB with a useless gadget to control speculative execution. Retpoline protection requires access to source code and recompilation.

**PoisonIvy:** PoisonIvy [39] is an architectural solution to track speculative data and prevent it from being exposed outside of the chip. The threat model focuses on accessing data while speculating on integrity verification. Their goal is different from ours: they seek to prevent the data from being speculatively read and therefore observed by a physical attacker that monitors the memory bus. PoisonIvy supports this capability by using information flow tracking to track data that is generated past a speculative check or data that is dependent on it. *SafeSpec* does not protect data from being speculatively exposed on the memory bus; rather it prevents side channel leakage. On the other hand, PoisonIvy does not prevent side channel leakage from speculatively accessed data. PoisonIvy results in approximately 20% slowdown in CPU performance.

Since the disclosure of the Spectre/Meltdown attacks, two closely relevant attacks have also been reported [6], [24]. Utilizing a verification tool, Trippel et.al. [24] discovered that by leveraging the invalidation message of cache coherence protocols, it is possible to replace Flush+Reload with Prime+Probe to retrieve the content fetched by speculative instructions. In the SgxPectre attack, Chen et.al. [6] demonstrated that it is possible to steal secret information from an SGX enclave using speculative execution.

## IX. Concluding Remarks

We presented a general principle for supporting speculative execution in a way that makes out-of-order processors immune to speculation-based attacks. The principle relies on leaving speculative state in shadow structures, and only committing this state once the instructions that generate them are guaranteed to commit. Thus, side-effects of misspeculation are hidden from the primary structures of the CPU, closing the vulnerability.

We applied the principle to protecting caches and TLBs of the CPU, which are the primary leakage vectors used in published speculation attacks. Our design completely closes all three published attacks, as well as new variants that we developed to leak through the I-cache or the TLBs. We showed that careful design is needed to prevent a form of leakage that can arise while instructions share the speculative state. We mitigate this leakage by sizing the speculative state conservatively. Constructed this way, transient attacks also become impractical.

The performance of the *SafeSpec* CPU was actually slightly higher than an unmodified CPU, despite conservative estimates on the shadow state. We explored some of the reasons for the differences in performance. First, the shadow state acts as an extension of the caches, raising their size, and potentially their bandwidth. Importantly, since misspeculated instructions do not commit, this leads to less clutter in the cache, although we lose out on some of the prefetching benefits of misspeculation.

We believe that the presented design represents a first step in many towards a principled protection of speculative execution. Our future work will consider these next steps that we discussed in Section VII.